# On the behaviour of spin-orbit connection of exoplanets

Bruno L. Canto Martins[1], Yuri S. Messias[1], Maria I. Arruda Gonçalves[1], Izan C. Leão[1], Roseane L. Gomes[1], Lorenza F. Barraza[1], Dasaev O. Fontinele[1], and José R. De Medeiros[1,*]

[1]Universidade Federal do Rio Grande do Norte, Departamento de Física Teórica e Experimental, Campus Central, 59072-970, Natal, RN, Brazil

[*] renan@fisica.ufrn.br

**Star–planet interactions play, among other things, a crucial role in planetary orbital configurations by circularizing orbits, aligning the star and planet spin and synchronizing stellar rotation with orbital motions. This is especially true for innermost giant planets, which can be schematized as binary systems with a very large mass ratio. Despite a few examples where spin–orbit synchronization has been obtained, there is no demographic study on synchronous regimes in those systems yet. Here we use a sample of 1,055 stars with innermost planet companions to show the existence of three observational loci of star–planet synchronization regimes. Two of them have dominant fractions of subsynchronous and supersynchronous star–planet systems, and a third less populated regime of potentially synchronized systems. No synchronous star–planet system with a period higher than 40 days has been detected yet. This landscape is different from eclipsing binary systems, most of which are synchronized. We suggest that planets in a stable asynchronous spin state belonging to star–planet systems in a supersynchronized regime offer the most favourable conditions for habitability.**

A short while ago, exoplanetary science celebrated the milestone of 5000 new worlds[1], mainly due to recent photometric space missions[2,3]. These discoveries reveal a remarkable planet zoo, with a large diversity of planetary system architectures[4,5] and physical parameters, including compact systems close to the hosting star[6,7]. This latter finding has inspired further studies into star-planet tidal interaction, as well as the search for relationships between stellar and planetary parameters, timescales, and implications for the planetary systems[8-13].

In this Article, we present a portrait of the synchronization states for the referred stars based on a refined sample of 1055 Kepler[2] and Transiting Exoplanet Survey Satellite (TESS)[3] main-sequence stars with confirmed planets or planet candidates[14-17]. The stellar sample definition, with the procedure for the main-sequence selection, as well as the planet selection, are presented in Methods. To analyze the synchronization regimes, we have used the ratio between the planet orbital period and the rotation period of the star, $P_{\rm orb}/P_{\rm rot}$, which corresponds to the measure of the synchronization degree for a given star-planet system. This ratio is equal to $\Omega_*/n$, where $\Omega_*$ is the rotational angular velocity of the star and $n$ is the mean orbital angular velocity. Synchronization occurs at $P_{\rm orb}/P_{\rm rot}=1.0$, whereas $P_{\rm orb}/P_{\rm rot}<1.0$ and $P_{\rm orb}/P_{\rm rot}>1.0$ define the subsynchronous and supersynchronous regimes, respectively.

## Results

Figure 1 shows the parameter $P_{\rm orb}/P_{\rm rot}$ as a function of the planet orbital period for the 1055 main-sequence stars composing our working sample. A remarkable portrait of the synchronization regimes emerges from that diagram, where we identify three subsamples, composed of synchronized ($P_{\rm orb}/P_{\rm rot}=1.0\pm0.1$), subsynchronized ($P_{\rm orb}/P_{\rm rot}<0.9$), and supersynchronized ($P_{\rm orb}/P_{\rm rot}>1.1$) star-planet systems. Indeed, these three subsamples delineate a continuous transition from subsynchronized to supersynchronized regimes, but the behavior of these subsamples depends on the considered orbital period range. For orbital periods shorter than about 2 days, one observes solely subsynchronous and synchronous star-planet systems. For orbital periods between 2 and 40 days, the three referred subsamples, subsynchronous, synchronous, and supersynchronous systems, are present. For orbital periods greater than 40 days, only supersynchronous systems are observed. To refine and quantify the transition between the above referred subsamples, we computed the 90th percentile of the period ratio $P_{orb}/P_{rot}$ distribution, from which it is possible to identify where the subsynchronous or supersynchronous systems are prevalent. Supplementary Fig. 3 illustrates the result of the 90th percentile. For orbital periods shorter than about 6.2 days, there is a prevalence of subsynchronous star-planet systems, although a small fraction of supersynchronous systems is present. For orbital periods between 6.2 and 13.5 days, we note a transition from subsynchronous to supersynchronous systems. For orbital periods larger than about 13.5 days, the occurrence of supersynchronous systems increases remarkably at the expense of synchronized star-planet systems, despite a small fraction of subsynchronous systems. However, the location of these subsamples, including that for the star-planet systems in a potential synchronization regime between 2 and 40 days, is particularly valid for the present working sample. The rotation period measurements from Kepler and TESS are biased to values shorter than about 45 days. In this sense, a fraction of stars may have rotation periods greater than the referred limit, partially fulfilling the lower right-hand region of Fig. 1 and possibly impacting the observed 40 days limit. However, according to different observational studies[18-20] and theoretical predictions[21-23], main-sequence stars with rotation periods greater than 45 days become scarce. As demonstrated in 'Sample bias effects' in Methods section, such scarcity is mainly a consequence of the anticorrelation between the spot modulation amplitude and the rotation periods observed for Sun-like stars, in such a way that rotation periods greater than 50-60 days become undetectable[20,24]. Within this limit, which corresponds to a semi-major axis less than about 0.3 au, the spin-orbit tidal coupling is expected to become efficient[25,26] enough for the occurrence of star-planet synchronized systems.

**The synchronization regimes of star-planet systems versus stellar binary systems**
Figure 1 also illustrates the distribution of the period ratio $P_{orb}/P_{rot}$ versus orbital period $P_{orb}$ for 816 Kepler eclipsing binaries (EBs) from a study on tidal synchronization in these systems[27]. From a direct comparison, the patterns in the $P_{orb}/P_{rot}$ versus $P_{orb}$ plane observed for star-planet systems differ from those observed for EBs. First, the referred study[27] shows that most synchronized EBs are concentrated in a range of orbital periods of less than 10 days and become scarce for periods greater than 30 days. In contrast, star-planet systems considered synchronized, namely those within the range $P_{orb}/P_{rot}=1.0\pm0.1$ in Fig. 1, are distributed for orbital periods up to 40 days with no noticeable concentration at a given $P_{orb}$ range. Second, the distribution of subsynchronized EBs is composed of a concentration of short-period EBs with $P_{orb}/P_{rot}$ about 13% below the synchronous region[27] and a remaining small fraction of subsynchronized EBs. Differently, for star-planet systems, there is a clear transition from subsynchronized to supersynchronized regimes with a significantly large fraction of subsynchronized systems. Finally, the distribution of the EBs in the supersynchronous region is relatively more populated than the subsynchronous region and less populated than the synchronous region. On the other hand, the distribution of supersynchronized star-planet systems is relatively less populated than the subsynchronous region, that is, the opposite of the observed for EBs, even if the overall distribution is rather similar to that observed for EBs. Another relevant view on the difference between the $P_{orb}/P_{rot}$ versus $P_{orb}$ behavior for star-planet systems and eclipsing binaries arises from the statistical distributions of the $P_{orb}/P_{rot}$ ratio for both samples. As observed in Supplementary Fig. 5, the $P_{orb}/P_{rot}$ distribution for star-planet systems spread substantially, whereas for eclipsing binaries, the distribution mainly concentrates around $P_{orb}/P_{rot}=1.0$. Such a difference is compatible with a weaker influence of the planetary tidal effects on the host star, contrasting with the binary systems.

**Discussion**
The portrait emerging from our results also provides a robust starting point for a qualitative evaluation of the occurrence of potentially asynchronous planets belonging to star-planet systems analyzed in the present study compared to what is predicted by theory. For that purpose, we have compared the star-planet systems obeying the different synchronization regimes (Fig. 1), with predictions of different planetary equilibrium spin states based on a global climate model[28]. Supplementary Fig. 6 shows an adapted stellar mass versus planetary semi-major axis diagram, stressing the separation of synchronous from asynchronous planets, orbiting low mass stars[28]. A clear scenario comes up from the referred diagram with a critical semi-major axis separating two regions. One region contains essentially supersynchronized star-planet systems located on the right side of the diagram. The other region, located from the center to the left side of the diagram, contains supersynchronized, synchronized, and subsynchronized systems, but the presence of subsynchronized systems dominates it. Interestingly, the region predicted by the models to be occupied by asynchronous rotation planets is essentially populated by supersynchronous star-planet systems, apparently independent of the planetary radius. This scenario has direct implications for the question of planetary habitability. Planets in systems tidally synchronized present additional difficulties for their habitability, even if they lie in the habitable zone. Because they have a perpetual darkness side, this aspect could represent an efficient cold-trap for water[28,29], strongly destabilize the carbonate-silicate cycle[30-32], and even produce atmospheric catastrophes in extreme cases[33,34]. Even though the asynchronous spin state is likely the result of complex interactions between stellar and atmospheric tides[35,36], Supplementary Fig. 6 points to an important fact. In principle, the best planets offering minor complex conditions for habitability are those in the stable asynchronous spin state belonging to star-planet systems in a supersynchronized regime, particularly those located inside habitable zones.

An additional feature emerging from the artwork presented in Fig. 1 that also brings attention is a lack of objects in the upper left corner, better defined in Supplementary Figure 4. Indeed, Sun-like main-sequence stars with rotation periods less than 0.1 days are not predicted by theoretical models[21-23], in agreement with different observational studies[18-20]. Also, according to theoretical predictions[37], a depletion of planets orbiting low-mass stars of about 0.6 $M_8$ to 1.0 $M_8$, with short rotation periods, typically $P_{rot}$<1.0 day, is expected for orbital periods shorter than about 2 to 5 days for planets with masses around 0.01 $M_J$ to 1.0 $M_J$, respectively. In addition, star-planet systems are scarce in the lower left corner, namely the region of systems with short orbital periods and larger rotational periods, for orbital periods shorter than about 1 day. Such a fact is also partly explained by theory[37], which predicts a deserted area of Jupiter mass planets around low-mass stars, with masses in the range of 0.6 $M_8$ to 1.0 $M_8$. Following theoretical studies[37-39], the rapid engulfment of planets and the efficient planet outward migration beyond the synchronization regime are possible explanations for the depleted area in the upper left corner. The scarcity of star-planet systems in the lower left corner would result from the hosting star spin-up due to planet migration[37].

Our results reveal the portrait of the behavior of synchronization regimes in star-planet systems, with two predominant fractions of subsynchronous and supersynchronous systems and a minor fraction of synchronized systems. This scenario contrasts with eclipsing binaries, for which synchronized systems dominate the synchronization regimes. Paralleling the Solar System bodies[40-42], we do not expect, despite a few examples[43-45], that measuring the rotation rate of exoplanets will be easy. Nevertheless, identifying the loci of star-planet synchronization regimes, associated with simple scaling arguments, can point the route to understanding how these worlds turn in the context of star-planet interaction and when asynchronous rotation is possible, which is an essential component for habitability studies[12,28,46].

## Methods
### The working sample

For the purpose of this study, we have composed a refined sample of main-sequence stars, with orbital periods and radius obtained from Kepler and TESS observations[47], as well as rotation periods computed from Kepler light curves[15] and TESS light curves[14]. The sample is composed of:

i. 934 Kepler Objects of Interest (KOI)[15];
ii. 121 TESS Objects of Interest (TESS)[14].

We applied a well-established procedure[48] for defining the main-sequence stage for the referred stars[49], which considers the stars' location in a Gaia color-magnitude diagram (Gaia CMD). This procedure makes it possible to refine the sample to avoid contamination from the subgiant and giant branches, as well as from binary stars.

Essentially, we have constructed a Gaia CMD to identify the location of the stars in the referred diagram. We selected Gaia magnitudes, color indexes, and parallaxes, considering the quality of the photometric solutions[48,50], taking only stars with $\sigma(G)/G$<0.01 and $\sigma(G_{RP})/G_{RP}$<0.01, where $G$, $G_{RP}$, $\sigma(G)$, and $\sigma(G_{RP})$ refer to the pass bands and the errors on the respective values used in the Gaia first and second data releases[50,51]. With these conditions, stars were placed in the Gaia CMD given in Supplementary Fig. 1. The figure also displays a 200 Myr isochrone with a metallicity of $[Fe/H]$=+0.25 from the Modules for Experiments in Stellar Astrophysics Isochrones & Stellar Tracks (MIST)[52-56], appropriately matching the main-sequence loci. We considered single main-sequence stars, those within 0.3 mag below and 0.9 mag above the 200 Myr isochrone, as shown in Supplementary Figure 1. The referred domain of magnitudes should encompass stars of different ages and metallicities, preventing contamination from the subgiant and giant branches, and by binary stars[48,57]. We list the entire sample with stellar and planetary parameters in Supplementary Table 1, available online in machine-readable form.

Planets were selected regardless of radii and masses. Nevertheless, we have considered only planets with $P_{orb}$ shorter than 10000 days, with a reported reference for the origin of their physical parameters. Furthermore, for those systems with multi-ple planets, we have considered the innermost companion.

**On the uncertainty of Kepler and TESS rotation periods**

This subsection discusses the level of uncertainty of Kepler and TESS rotation period measurements to control the effects of possible biases, particularly on the scenario emerging from Fig. 1. Supplementary Fig. 2 displays the distributions of errors on the $P_{orb}/P_{rot}$ ratio, $e(P_{orb}/P_{rot})$, for Kepler and TESS rotation periods measurements, represented by the blue and orange colors, respectively.

We have estimated a typical error of 0.006 for the absolute ratio $P_{orb}/P_{rot}$ considering the whole sample. We also estimated that 87% of the $P_{orb}/P_{rot}$ values have errors less than 0.1. Considering the $P_{orb}/P_{rot}$ values range, these errors represent negligible impact in changing the fractions of synchronous, subsynchronous, and supersynchronous star-planet systems. In addition, we performed a Monte Carlo test by fluctuating the $P_{orb}/P_{rot}$ values within the individual errors and computed the average amount of stars that could change the synchronization regime. From this test, we estimated that:

I. 16% of the synchronized star-planet systems could have a different regime (super- or subsynchronous);
II. 0.2% of the supersynchronous systems could be synchronized or subsynchronous;
III. 0.06% of the subsynchronous systems could be synchronized or supersynchronized.

Therefore, the overall results emerging from Fig. 1 are very stable relative to the uncertainty in the $P_{orb}/P_{rot}$ measurements. The most affected region would naturally be that of the synchronized systems, distributed in a narrow region compared to the other two regimes but still with a proper fraction for keeping our main conclusions.

**A Transition from subsynchronous to supersynchronous regime**

Following the procedure presented in a recent study[27], we dynamically computed a 90th percentile test of the period ratio distribution to check the apparent transition from subsynchronous to the supersynchronous regimes observed in Fig. 1. For each star-planet system, we take the other 29 star-planet systems with the nearest orbital periods to calculate the percentile. Indeed, we have computed the percentile for different box sizes obtaining a similar behavior. Supplementary Fig. 3 shows the results of the referred percentile test.

**Sample bias effects**

We show in the following that bias effects in the working sample have no significant consequence in the results presented in Fig. 1. First, the sample is subject to a selection bias associated with spot modulation detectability which has a direct impact on rotation period measurements[20,24,58,59], for rotation periods longer than 50 to 60 days, the frequency of period detectability decreases dramatically, a fact also predicted by theoretical models[21]. In addition, the relation between the rotation period and the amplitude of the rotational modulation points for an anti-correlation where the rotation signal becomes almost undetectable at periods longer than 50-60 days[20].

Supplementary Fig. 4 presents the same distribution as Fig. 1 with specific limits for the potential detectability of the rotation periods. The $P_{rot}$ values shorter than 0.1 days are not expected to occur based on observational studies[18,19,60] and theoretical predictions[18,22,23]. For $P_{rot}$ greater than 60 days, the amplitude of rotational modulation decreases considerably[20,24], thus minimizing the occurrence of star-planet systems.

**The distribution of $P_{orb}/P_{rot}$ for star-planet systems and eclipsing binaries**

This subsection presents the statistical distribution of the $P_{orb}/P_{rot}$ for the working sample of 1055 KOI and TOI star-planet systems and for 764 Kepler eclipsing binaries (EBs). Here, we are considering only the binary systems defined as the EB main population from the sample of 816 Kepler systems[27] described in the main text.

**Asynchronism among the present working sample of star-planet systems**

The asynchronism of planets is a fundamental aspect of the conditions of habitability[28]. In this context, the present study offers a unique possibility to compare the loci of the star-planet synchronization regimes emerging from the present analysis, with theoretical predictions, particularly on the location of asynchronous planets relative to habitable zones. To that end, we have adapted the scenario for stellar mass distribution versus planetary semi-major axis from theoretical predictions based on a global climate model[28]. Such a model is used to simulate planets in the habitable zone of low-mass stars, with different atmospheric masses, compositions, and incoming stellar fluxes, from which it is possible to quantify the torque induced by thermal tides. Consequently, these models offer the possibility of identifying different planet synchronization regimes[28].

Supplementary Fig. 6 shows the stellar mass distribution as a function of the planetary semi-major axis for our sample of 1055 main sequence stars, with both parameters taken from the NASA Exoplanet Archive[47]. This figure shows the location of planets' synchronous and asynchronous regimes based on a model for thin atmospheres (1 bar pressure) and thick atmospheres (10 bars pressure)[28]. Indeed, these models demonstrate that asynchronism mediated by thermal tides should affect a considerable fraction of planets in the habitable zone of lower-mass stars.

**Data Availability**
All data generated or analyzed during this study are included in this published article (and its Supplementary Information files).

**Acknowledgements**
Research activities of the board of observational astronomy at the Federal University of Rio Grande do Norte are supported by continuous grants from the Brazilian funding agencies CNPq and FAPERN. The present project was granted by the CNPq (Universal/409345/2021-0) and CAPES (PRINT/88881.310208/2018-01). This study was financed in part by the Coordenação de Aperfeiçoamento de Pessoal de Nível Superior - Brasil (CAPES) - Finance Code 001. Y.S.M., M.I.A.G., R.L.G., L.F.B., and D.O.F. acknowledge CAPES graduate fellowships. CNPq research fellowship are acknowledge by B.L.C.M., I.C.L., and J.R.M. This work includes data collected by the TESS and Kepler missions. This research has made use of the NASA Exoplanet Archive, which is operated by the California Institute of Technology, under contract with the National Aeronautics and Space Administration under the Exoplanet Exploration Program.

**Author Contributions Statement**
B.L.C.M. and J.R.M. led the project and wrote the manuscript. Y.S.M. and I.C.L. led the data analysis and art preparation. I.C.L., R.L.G., L.F.B., and D.O.F. led software preparation and data reduction. M.I.S.G. revised the manuscript. All authors discussed and commented on the manuscript.

**Competing Interests Statement**
The authors declare no competing interests.

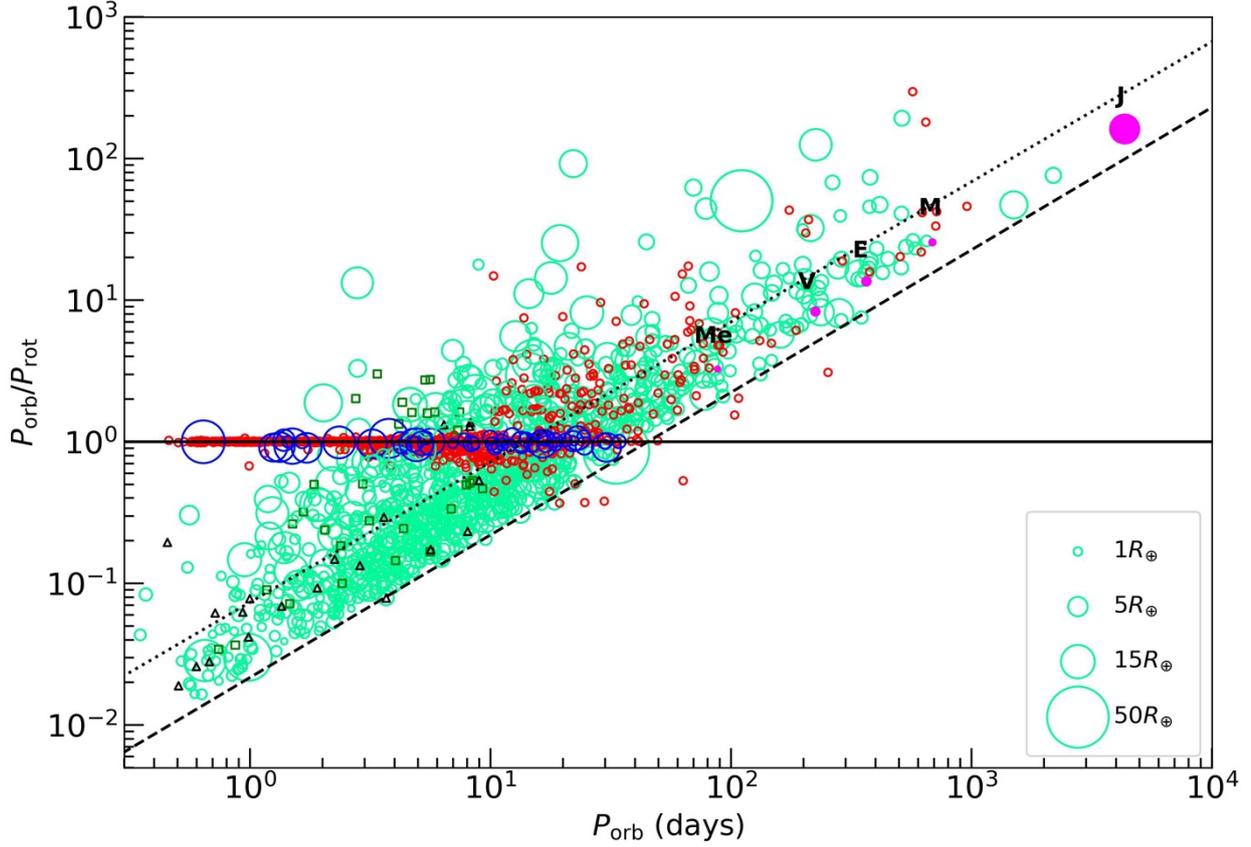

**Figure 1: The distribution of the period ratio $P_{orb}/P_{rot}$ versus the orbital period $P_{orb}$.** This figure shows our sample of 1055 main-sequence stars with confirmed planets or planet candidates[14-17], represented by blue and soft green open circles. Open blue circles represent the potentially synchronized systems, here assumed as those systems with $P_{orb}/P_{rot}$=1.0±0.1. In addition, the figure also displays a sample of 816 Kepler eclipsing binaries (EB)[27], with red open circles, dark green open squares, and black open triangles representing EB main population, asynchronous short-period EBs, and possible exoplanets or brown dwarfs, respectively. We also illustrate Solar System planets Mercury (Me), Venus (V), Earth (E), Mars (M), and Jupiter (J) as solid magenta circles. Circle sizes are proportional to the planet's radius squared. The black horizontal solid line corresponds to synchronization at $P_{orb}/P_{rot}$=1.0. The diagonal dotted and dashed lines indicate a conservative rotation period limit of 15 and 45 days for TESS and Kepler, respectively.

**Supplementary Information**

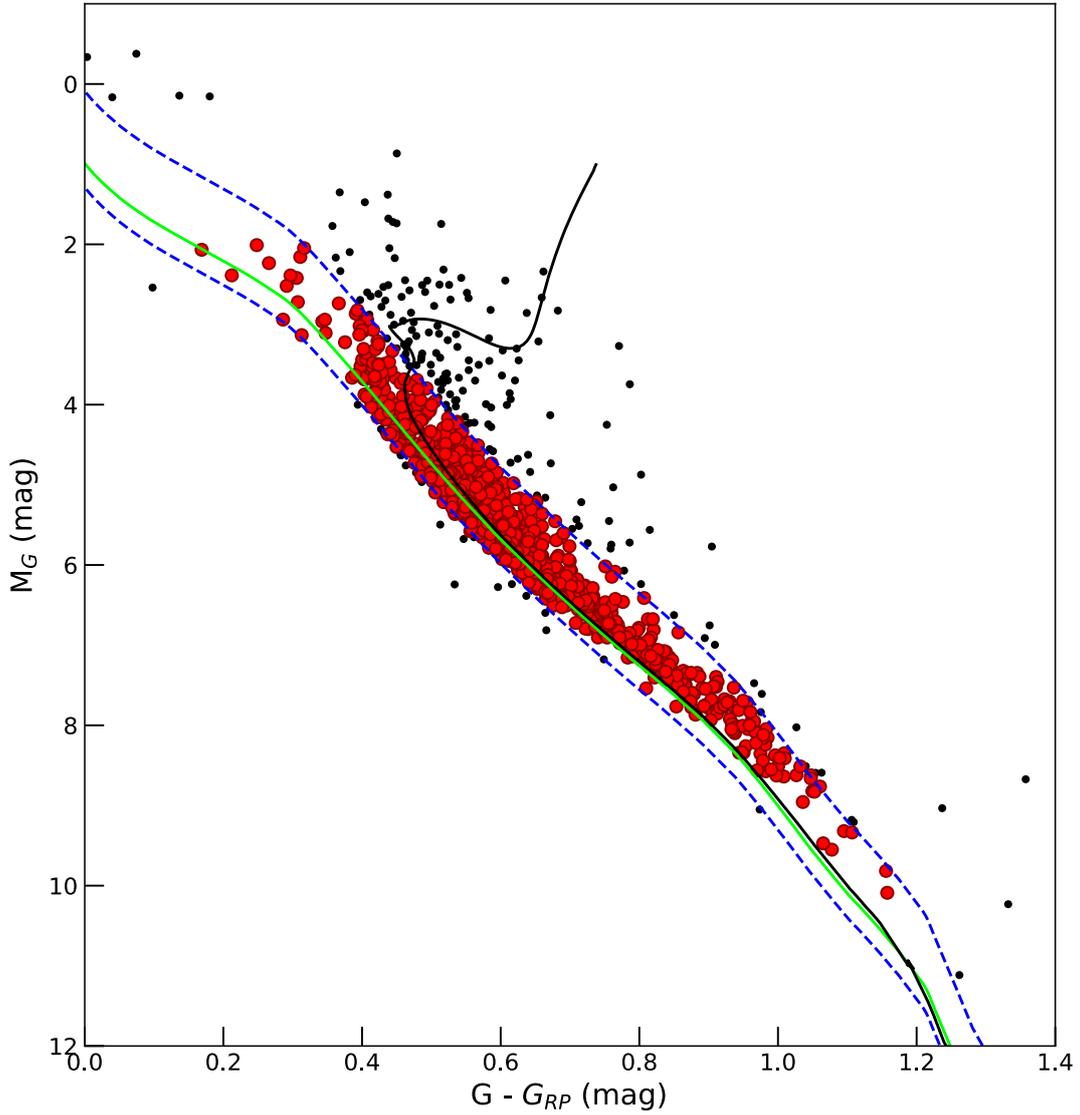

**Supplementary Figure 1: Gaia color–magnitude diagram for the initial sample of KOIs and TOIs stars.** Red dots represent the 1055 main-sequence stars, namely those filling the Gaia photometric and parallax criteria described in the text. The solid green curve represents the MIST isochrone[1,2,3,4,5] with an age of 200 Myr to identify the main sequence; the solid black curve represents a MIST isochrone with an age of 4.5 Gyr to illustrate the subgiant and giant branches. The dashed blue curves represent the 200 Myr isochrone shifted down and up by 0.3 and 0.9 mag, respectively, defining the extent of the main-sequence stars.

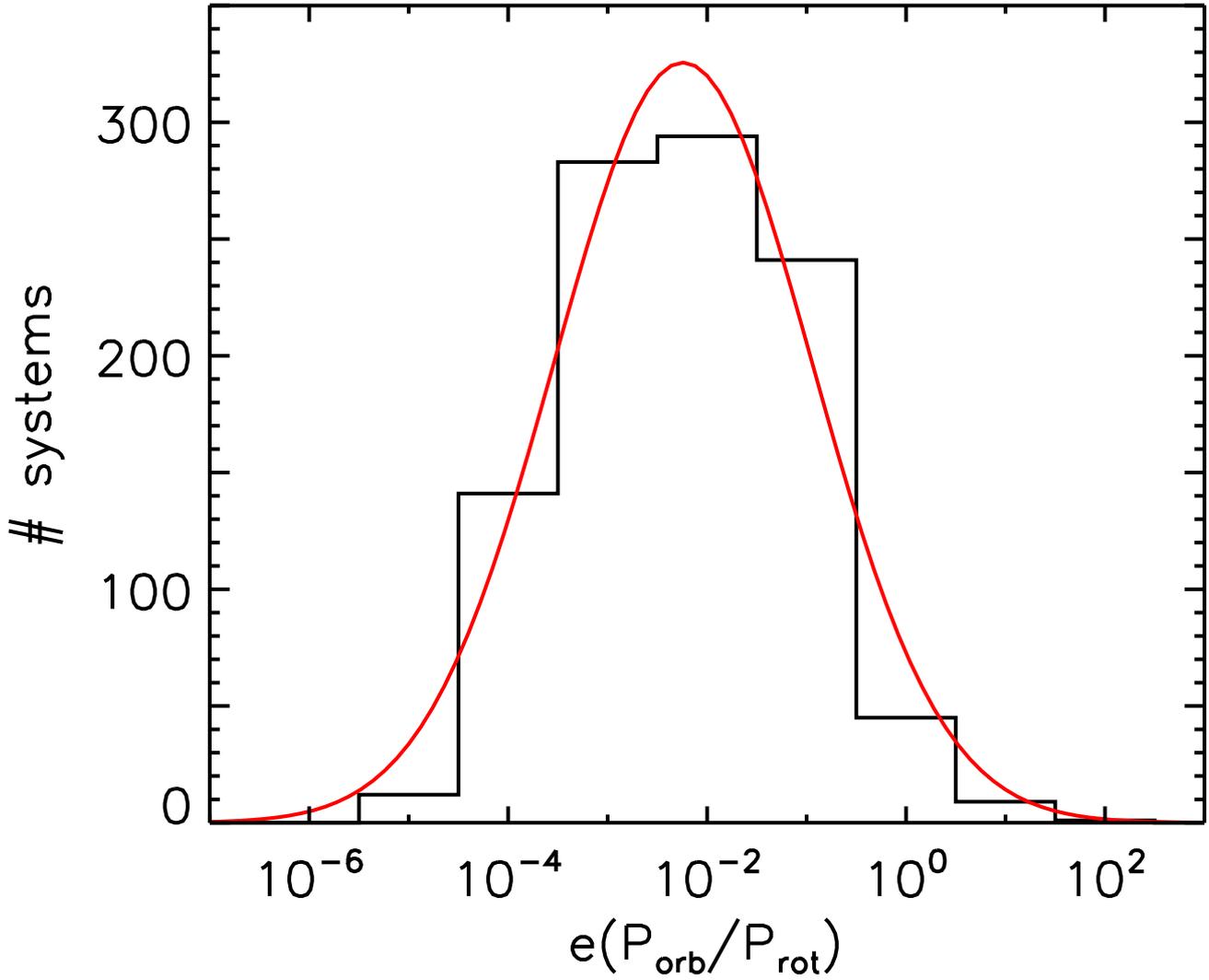

**Supplementary Figure 2: Distribution of the uncertainties on the $P_{orb}/P_{rot}$ ratio, $e(P_{orb}/P_{rot})$, of the star-planet systems studied in Figure 1.** The black histogram represents the observational data, while the solid red curve is a gaussian fit with the data for the uncertainties given in a logarithmic scale. The $e(P_{orb}/P_{rot})$ value corresponding to the gaussian maximum gives the typical uncertainty of the whole sample.

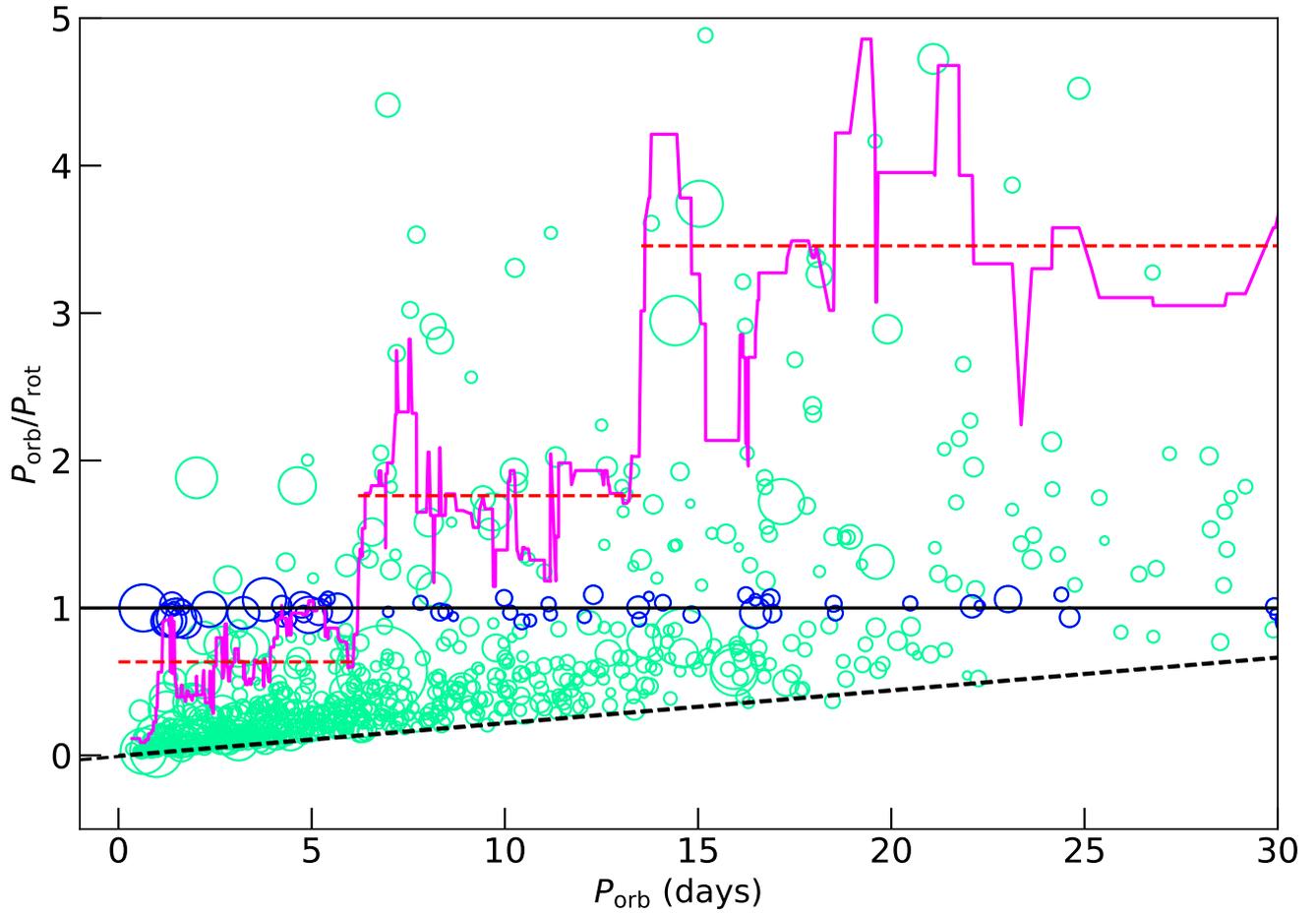

**Supplementary Figure 3: A zoom of the distribution of $P_{orb}/P_{rot}$ vs. $P_{orb}$ for the present working sample of star-planet systems, presented in Figure 1.** The solid magenta curve indicates the running 90th percentile computed, with the dashed red lines indicating the mean value of the 90th percentile for $P_{orb} \leq 6.2$ days, $6.2 < P_{orb} \leq 13.5$ days, and $P_{orb} > 13.5$ days. The black horizontal line corresponds to synchronization at $P_{orb}/P_{rot}=1.0$. The black dashed diagonal line corresponds to a conservative rotation-period limit of 45 days, from which measuring longer period signals becomes difficult[6,7]. Circle sizes are proportional to the planet's radius squared.

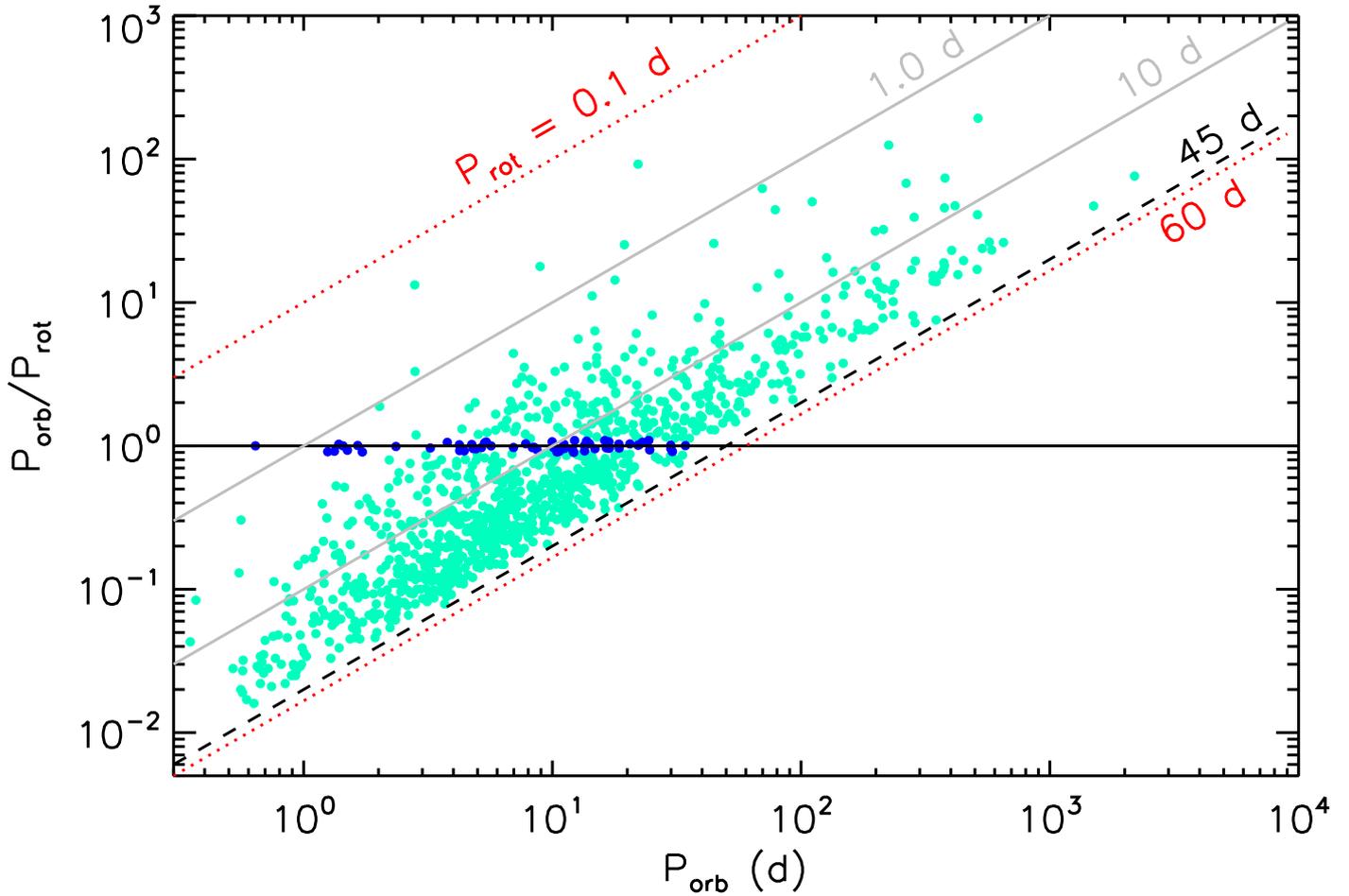

**Supplementary Figure 4: Rotation period detectability limits in the distribution of $P_{orb}/P_{rot}$ versus $P_{orb}$.** Green and blue circles depict the sample of 1055 KOI and TOI star-planet systems shown in Figure 1. The downer dotted red diagonal line represents the expected physical limit of about 60 days for detecting rotational modulation, whereas the dashed black diagonal line of 45 days indicates the maximum detectability period for the present working sample. The upper dotted red diagonal line illustrates the limit of 0.1 days below which the occurrence of stars is not expected from observations and theoretical models. The solid gray diagonal lines delimit a scale for 1.0 and 10 days rotation periods. The black horizontal line corresponds to synchronization at $P_{orb}/P_{rot}=1.0$.

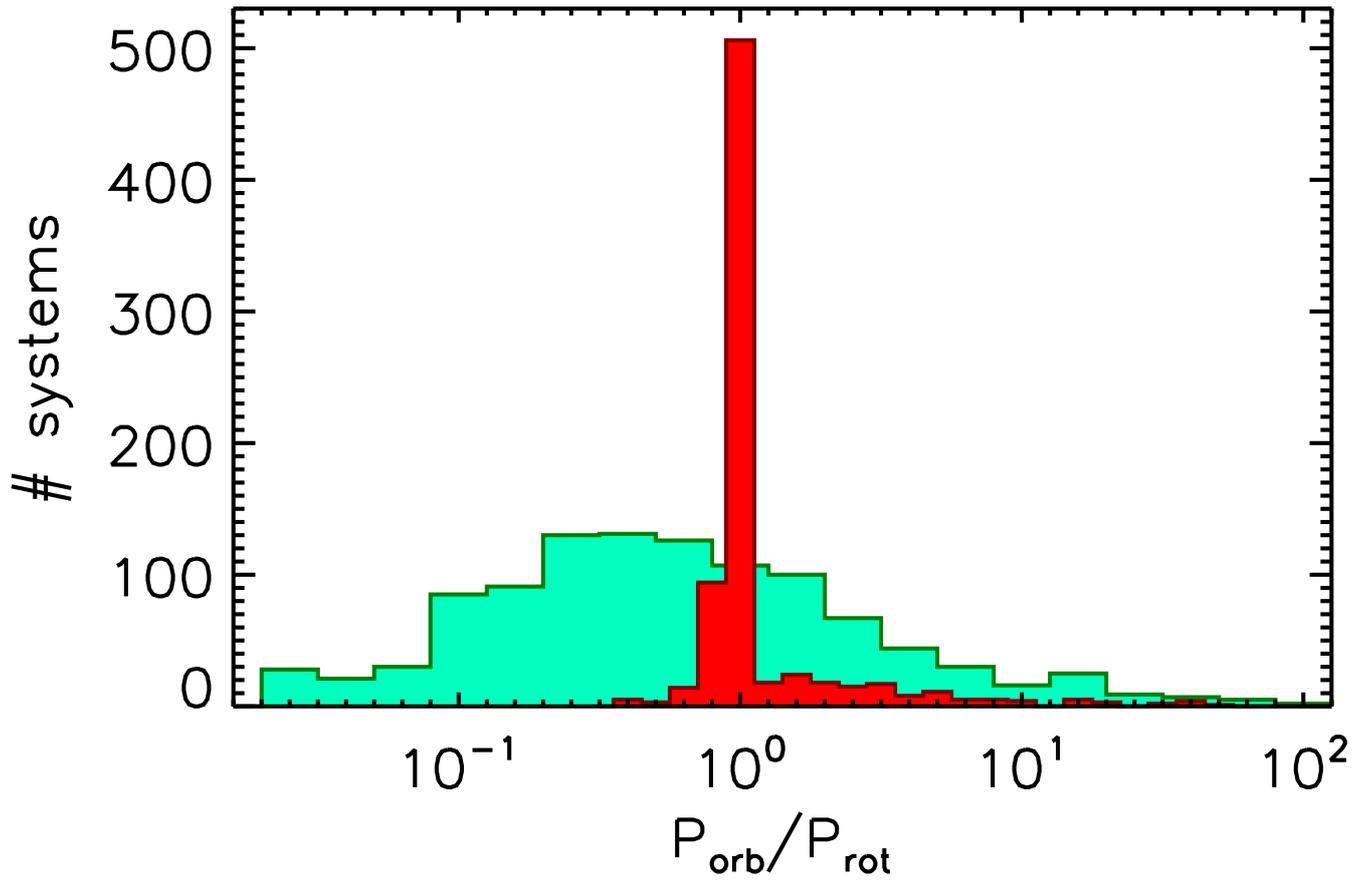

**Supplementary Figure 5: The statistical distributions of Porb/Prot for star-planet systems and eclipsing binaries.** The green color refers to the sample of 1055 KOI and TOI star-planet systems, whereas the red color represents a sample of 764 Kepler eclipsing binaries systems[6]. These distributions are overplotted for comparison.

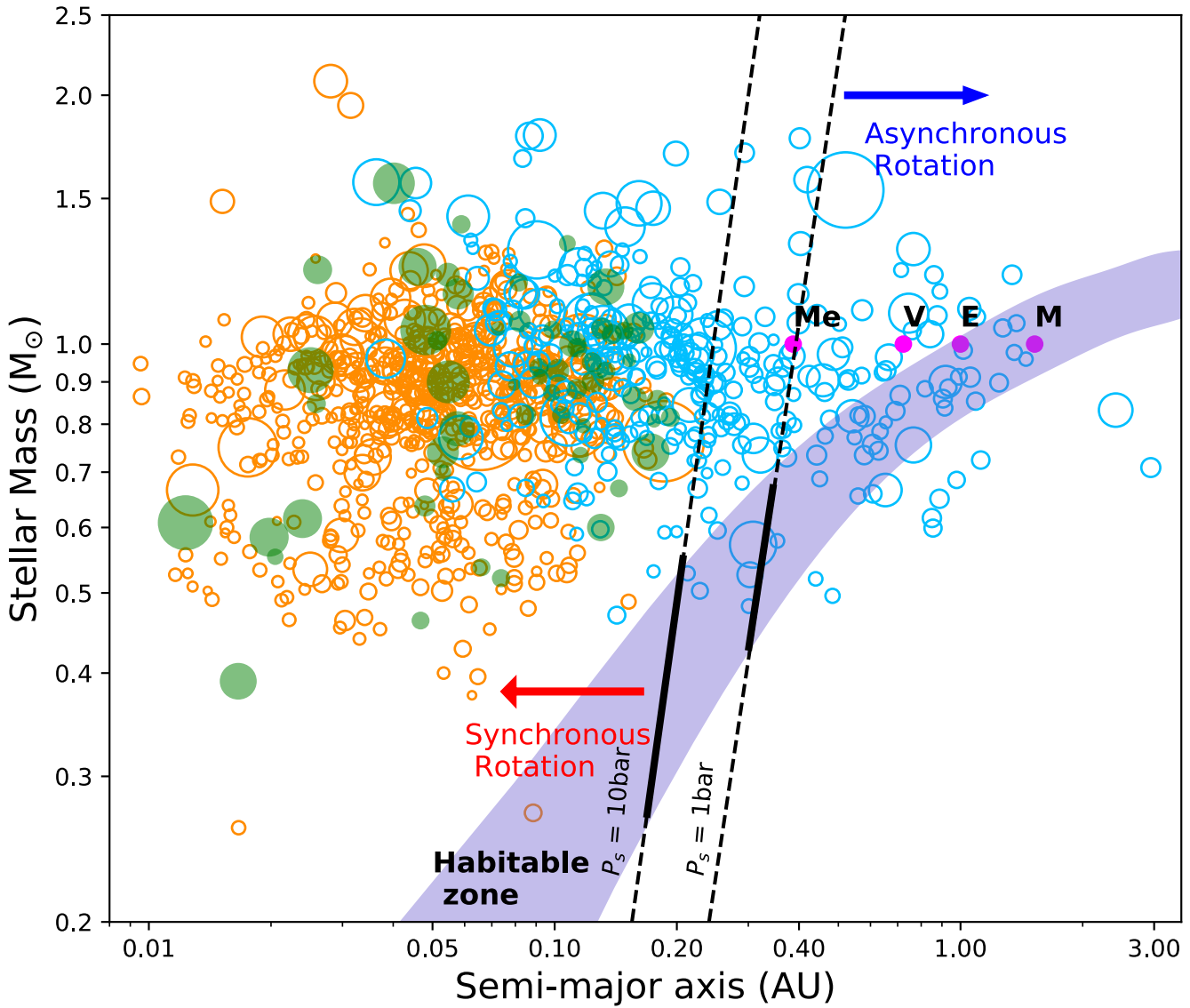

**Supplementary Figure 6: Stellar mass versus semi-major axis for the present working sample.** Subsynchronous and supersynchronous systems are pictured by orange and blue circles, respectively. Green circles represent the synchronized systems. Circle sizes are proportional to the planet's radius squared. Solid lines indicate the critical semi-major axis separating synchronous from asynchronous planets inside the habitable zones (blue strip), whereas dashed lines represent an extrapolation outside the habitable zone for two different models[8]. Solar planets (Mercury, Venus, Earth, and Mars) are also pictured.